\documentclass[prd,superscriptaddress,floatfix,amsmath,footinbib,amssymb,twocolumn]{revtex4}
\usepackage{amssymb}
\usepackage{amsmath}
\usepackage{amsfonts}
\usepackage{tikz}
\usepackage{bm}

\usepackage{epsfig}
\usepackage{t1enc}
\usepackage{soul}
\usepackage{color}

\usepackage{mathrsfs}
\usepackage{url}
\usepackage[all]{xy}
\usetikzlibrary{calc,patterns,angles,quotes}

\begin{document}

\title{Reply to Galilean invariance without superluminal particles}

\author{Andrzej Dragan}
\affiliation{Institute of Theoretical Physics, University of Warsaw, Pasteura 5, 02-093 Warsaw, Poland}
\affiliation{Centre for Quantum Technologies, National University of Singapore, 3 Science Drive 2, 117543 Singapore, Singapore}

\author{Artur Ekert}
\affiliation{Centre for Quantum Technologies, National University of Singapore, 3 Science Drive 2, 117543 Singapore, Singapore}
\affiliation{Mathematical Institute, University of Oxford, Woodstock Road, Oxford OX2 6GG, United Kingdom}

\maketitle

In a recent paper \cite{Dragan2020} we have analyzed consequences of the Galilean principle of relativity stating that motion with {\em any} constant speed does not change the laws of physics. We derived all possible reference frame transformations and found that only two families of observers are mathematically allowed: the subluminal observers characterized by the usual Lorentz transformation:
\begin{eqnarray}
\label{Lorentz}
x' &=& \frac{x-Vt}{{\sqrt{1-V^2/c^2}}},\nonumber \\
t' &=& \frac{t-Vx/c^2}{{\sqrt{1-V^2/c^2}}}
\end{eqnarray}
for $V<c$ and the superluminal ones given by:
\begin{eqnarray}
\label{SuperLorentz}
x' &=& -\frac{V}{|V|}\frac{x-Vt}{{\sqrt{V^2/c^2-1}}},\nonumber \\
t' &=& -\frac{V}{|V|}\frac{t-Vx/c^2}{{\sqrt{V^2/c^2-1}}}
\end{eqnarray}
for $V>c$. It is usual to discard the mathematically valid solutions \eqref{SuperLorentz} as "unphysical", which, to paraphrase Orwell, is to say that "all inertial frames of reference are equal, but some of them are more equal than the others". In our work \cite{Dragan2020} we have shown that taking into account both families of observers on equal footing, leads to a familiar non-deterministic quantum-mechanical picture of reality involving quantum superpositions and complex probability amplitudes. We have concluded that basic laws of quantum theory are a straightforward consequence of Einsteinian relativity extended to superluminal observers.

In a recent paper \cite{Grudka2021} Grudka and W\'ojcik question validity of some of our results and propose an alternative interpretation of the equations. The first criticism is based on their claim that the solutions \eqref{SuperLorentz} can be ruled out based on a symmetry argument involving frames with reversed space coordinates. We would like to point out that since both solutions \eqref{Lorentz} and \eqref{SuperLorentz} equally satisfy the Galilean principle of relativity and equally preserve the constancy of the speed of light, ruling any of them out must involve some additional limiting assumption. Indeed, in \cite{Grudka2021} that assumption is present and it is rooted in the insufficiently general intuition taken from subluminal reference frames that flipping the spatial coordinates of both frames of reference should be a symmetry. This extra assumption is not true for motions with superluminal speeds, since the superluminal branch \eqref{SuperLorentz} is a hyperbolic rotation of spacetime by the angle larger than $\frac{\pi}{4}$. Therefore reversing the space coordinate of the resting observer should be equivalent of reversing the temporal coordinate of the superluminally moving observer. This is seen in \eqref{SuperLorentz} and especially in the limiting case of motion with infinite speed, for which the spacetime axes are both rotated by $\frac{\pi}{2}$ and \eqref{SuperLorentz} reduces to $x' = ct$ and $ct' = x$.

Next, Grudka and W\'ojcik point out that conclusions of \cite{Dragan2020} are based on a "hidden assumption" that there exist superluminal observers. In fact, this assumption is not hidden at all, it is just a part of the Galilean principle of relativity, which puts no restriction on possible velocities of inertial observers whatsoever. Quite contrary, it takes an extra assumption to rule out the superluminal solutions from special relativity as "unphysical" (and it remains a mystery, how is it not contradicting the principle relativity). Grudka and W\'ojcik introduce, what they call "a more natural interpretation" of the coordinate system given by \eqref{SuperLorentz}, in which they reinterpret $ct'$ as the spatial coordinate and $x'$ as the temporal coordinate. Contrary to the suggestion of the authors of \cite{Grudka2021} this choice is not a mater of interpretation or convention. A temporal coordinate of any observer, by definition, must be parallel to the trajectory of that observer, so that she is resting in her own co-moving frame. Therefore the transformation to the rest frame of the superluminal observer must be given by \eqref{SuperLorentz} with $ct'$ being her temporal coordinate and $x'$ her spatial coordinate. No room for interpretation is left here and therefore being superluminal is relative.

Grudka and W\'ojcik also make an interesting final remark concerning the Schwarzschild spacetime. Schwarzschild coordinates characterizing such a spacetime correspond to stationary observers placed at fixed distances from the event horizon. Such observers can be subluminal only above the horizon, and under the horizon they require superluminal motions. The sign flip in the metric therefore signifies the transition from a subluminal to a superluminal family of stationary observers residing under the event horizon in a fixed distance from that horizon. Therefore, the Galilean principle of relativity can be castrated from half of the solutions resulting in conventional relativity, but this is no longer possible in curved spacetimes.


\begin{thebibliography}{9}

\bibitem{Dragan2020}
A. Dragan and A. Ekert, New J. of Phys. {\bf 22}, 033038 (2020).

\bibitem{Grudka2021}
A. Grudka and A. Wójcik, arXiv:2112.05658 [quant-ph] (2021).

\end{thebibliography}
\end{document}